\documentclass[amsmath, amssymb, amsfonts, twocolumn, footinbib, superscriptaddress]{revtex4-1}
\usepackage{graphicx}
\usepackage{graphics}
\usepackage{dcolumn}
\usepackage{bm}
\usepackage{amssymb}
\usepackage{amsmath}
\usepackage{amsfonts}
\usepackage{epsfig}
\usepackage{mathbbol}
\usepackage{latexsym}
\usepackage{dcolumn}
\usepackage{hyperref}
\usepackage{float}
\usepackage[normalem]{ulem}
\usepackage[usenames, dvipsnames]{color}
\usepackage{epstopdf}

\hypersetup{                    
    colorlinks,
    citecolor=blue,
    filecolor=blue,
    linkcolor=blue,
    urlcolor=blue
}

\newcommand {\ket} [1] {| #1 \rangle}
\newcommand {\bkt} [1] {\langle #1 \rangle}

 \newcommand {\beq}{\begin{equation}}
\newcommand {\eeq}{\end{equation}}
\newcommand {\beqn}{\begin{eqnarray}}
\newcommand {\eeqn}{\end{eqnarray}}
\newcommand {\bit}{\begin{itemize}}
\newcommand {\eit}{\end{itemize}}

\newcommand{\dps}{\displaystyle}

\newcommand{\om}{\iffalse}

\definecolor{mygray}{gray}{0.6}
\definecolor{gold}{RGB}{150, 150, 10}
\definecolor{mygreen}{RGB}{40, 200, 100}

\newcommand{\bw}[1]{{\textcolor{blue}{#1}}}

\begin{document}
\title{Signatures of quantum mechanical Zeeman effect in classical transport due to topological properties of two-dimensional spin$-3/2$ holes}

\author{E. Marcellina}
\altaffiliation{Present address: School of Physical and Mathematical Sciences, Nanyang Technological University, 21 Nanyang Link, Singapore 637371. Email: emarcellina@ntu.edu.sg}
\affiliation{School of Physics and Australian Research Council Centre of Excellence in Future Low-Energy Electronics Technologies, The University of New South Wales, Sydney 2052, Australia}

\author{Pankaj Bhalla}
\affiliation{Beijing Computational Science Research Center, Beijing 100193, China}

\author{A.~R.~Hamilton}
\affiliation{School of Physics and Australian Research Council Centre of Excellence in Future Low-Energy Electronics Technologies, The University of New South Wales, Sydney 2052, Australia}

\author{Dimitrie Culcer}
\affiliation{School of Physics and Australian Research Council Centre of Excellence in Future Low-Energy Electronics Technologies, The University of New South Wales, Sydney 2052, Australia}

\begin{abstract}
The Zeeman interaction is a quantum mechanical effect that underpins spin-based quantum devices such as spin qubits. Typically, identification of the Zeeman interaction needs a large \textit{out-of-plane} magnetic field coupled with ultralow temperatures, which limits the practicality of spin-based devices. However, in two-dimensional (2D) semiconductor holes, the strong spin-orbit interaction causes the Zeeman interaction to couple the spin, the magnetic field, and the momentum, and has terms with different winding numbers. In this work, we demonstrate a physical mechanism by which the Zeeman terms can be detected in classical transport. The effect we predict is very strong, and tunable by means of both the density and the in-plane magnetic field. It is a direct signature of the topological properties of the 2D hole system, and a manifestation in classical transport of an effect stemming from relativistic quantum mechanics. We discuss experimental observation and implications for quantum technologies. 
\end{abstract}
\date{\today}
\maketitle

The recent years have witnessed rapid development of materials and structures in which relativistic quantum mechanical effects, such as the spin-orbit interaction, play a dominant role in future spin-based device concepts \cite{Zhang_PhyTod2010, Kane_RMP2010, Kormanyos_2DM_2015, Armitage_Weyl_RMP_2018}. A natural place where such effects are important is state-of-the-art semiconductor hole devices \cite{Manfra-2005-APL, Habib2009, Hao-2010-NL, Chesi-2011-PRL, Daisy-2016-NL, Srinivasan-2017-PRL, Korn-2010-NJP, Watson-2011-PRB, Srinivasan-2016-RPB, Papadakis-Shayegan-2000-PRL, Shayegan-2002-PRB, Joost-2014-NL, Nichele-Ensslin-2014-PRB, Ota-Tarucha-2004-PRL, Ono-Tarucha-2002-Sci, Tarucha-2007-RMP, Croxall2013, Katsaros-2016-NL, Gerl2005, Clarke2007}. The impressive progress in this field over this decade is motivated by the intense interest in hole systems as building blocks in quantum computing architectures and as the next generation of nanoscale topological devices \cite{Cuan-2015-EPL, Biswas-2015-EP, Zwanenburg-2013-RMP, Conesa-Boj-2017-NL, Matthias-Zwanenburg-2016-APL, Brauns-Zwanenburg-2016-PRB,Mueller-Zwanenburg-2015-NL, Fanming-Kouwenhoven-2016-NL,Maurand2016,Jo-2017-PRB, Watzinger2018, Liles2018}. The strong spin-orbit coupling in holes \cite{Roland, Moriya-2014-PRL, Tutul-2014-AP, Shanavas-2016-PRB, Akhgar-2016-NL,Pang-JPCM-2010, HongMarcellina2018, Marcellina2018} together with weak hyperfine interactions enable low-power electrical spin manipulation \cite{Bulaev-2005-PRL,Nichele-Kouwenhoven-2017-PRL} and long coherence times \cite{Korn-2010-NJP, Salfi-2016-PRL, Salfi-2016-Nanotechnology,Petta-Charlie-Marcus-2005-Sci}. Furthermore,  strongly spin-orbit coupled systems can exhibit proximity-induced superconductivity \cite{Hendrickx2018}, which can open a pathway towards topological quantum computation via Majorana bound states with non-Abelian statistics \cite{Lutchyn-2010-PRL, Alicea-2011-NP, Gill-2016-APL, Alestin-India-arXiv, Roberto-PRB-2001, Liang2017}. However, a challenge that can limit the practicality of spin-based devices is that the detection of spin states often requires a large magnetic field and/or ultralow temperatures \cite{Habib2009,Hao-2010-NL,Korn-2010-NJP,Srinivasan-2016-RPB, Papadakis-Shayegan-2000-PRL, Shayegan-2002-PRB,Ono-Tarucha-2002-Sci,Ota-Tarucha-2004-PRL,Chesi-2011-PRL,Joost-2014-NL,Nichele-Ensslin-2014-PRB,Daisy-2016-NL,Katsaros-2016-NL,Srinivasan-2017-PRL,Miserev2017a,Marcellina2018,Watzinger2018,Liles2018,Hendrickx2018}. In light of this it is natural to investigate the signatures of spin in classical transport. 

\begin{figure}
	\centering
	\includegraphics[scale=0.78]{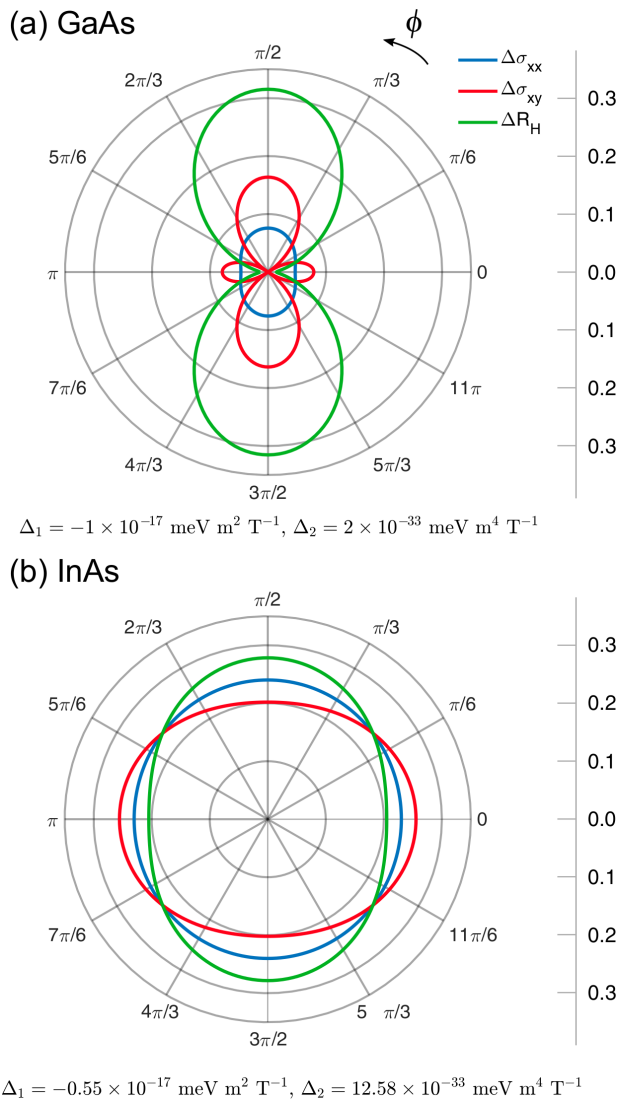}
	\caption{The correction for the longitudinal conductivity $\Delta \sigma_{xx}$, transverse conductivity $\Delta \sigma_{xy}$, and Hall coefficient $\Delta R_{\mathrm{H}}$ for (a) GaAs and (b) InAs. We define $\Delta R \equiv |R - R_0|/R_0$ where $R_0$ is the magnetoresistance or Hall resistance at zero magnetic field. Here  $k_F = 1\times10^{8}$ m$^{-1}$ and $B_{\parallel}$ = 1 T.}
	\label{fig:Results}
\end{figure}

In this work we examine the signatures of the Zeeman interaction, which is fundamental to spin qubit dynamics, spin-based quantum information processing, and Majorana states in semiconductor-superconductor hybrids, in classical transport. We find that, in strongly spin-orbit coupled systems, the Zeeman interaction has a significant and measurable effect on the classical magnetoresistance. Our model system is a two-dimensional hole gas (2DHG) in a symmetric quantum well, subject to a weak out-of-plane magnetic field $B_z$ (of order milliTeslas) and a sizable in-plane magnetic field $(B_x, B_y)$ (of order a few Teslas), where the Rashba spin-orbit interaction \cite{Rashba1984} is absent. As Fig.~\ref{fig:Results} shows, the Zeeman interaction with an in-plane magnetic field gives rise to a sizable anisotropy in both the longitudinal conductivity and the Hall coefficient, implying, in particular, that the Hall coefficient in this case is not simply a measure of the carrier density and is strongly influenced by the Zeeman interaction. Fundamentally this effect occurs because, unlike spin-1/2 electrons, holes are described by an effective spin-3/2. This means that there are four possible projections onto the spin quantization axis (i.e. the axis perpendicular to the interface, denoted here by $z$). The projections $m_J = \pm 3/2$ ($m_J = \pm 1/2$) have a heavier (lighter) effective mass, thus these states are termed as heavy (light) holes. Since, in two dimensions, the heavy holes projection is parallel to the $z-$axis, they exhibit a peculiar Zeeman interaction with an in-plane magnetic field $B_\parallel$, which couples the spin simultaneously to the external magnetic field and to the momentum $k$. It has long been known that a term $\propto B_\parallel k^2$ exists quadratic in the wave vector ${\bm k}$ \cite{Roland}, while recently a new term  $\propto B_\parallel k^4$ was predicted \cite{Miserev2017} and experimentally identified \cite{Miserev2017a}. These two terms have different winding directions and make large contributions to the velocity operator and hence to the current. The higher-order coupling between $B_{\parallel}$ and $k$ also manifests in an extremely density-dependent in-plane $g$-factor, whereby we have demonstrated that the $g-$factor became enhanced by 300\% as the density was tripled \cite{Marcellina2018}. 

\begin{figure}
	\centering
	\includegraphics[scale=1.05]{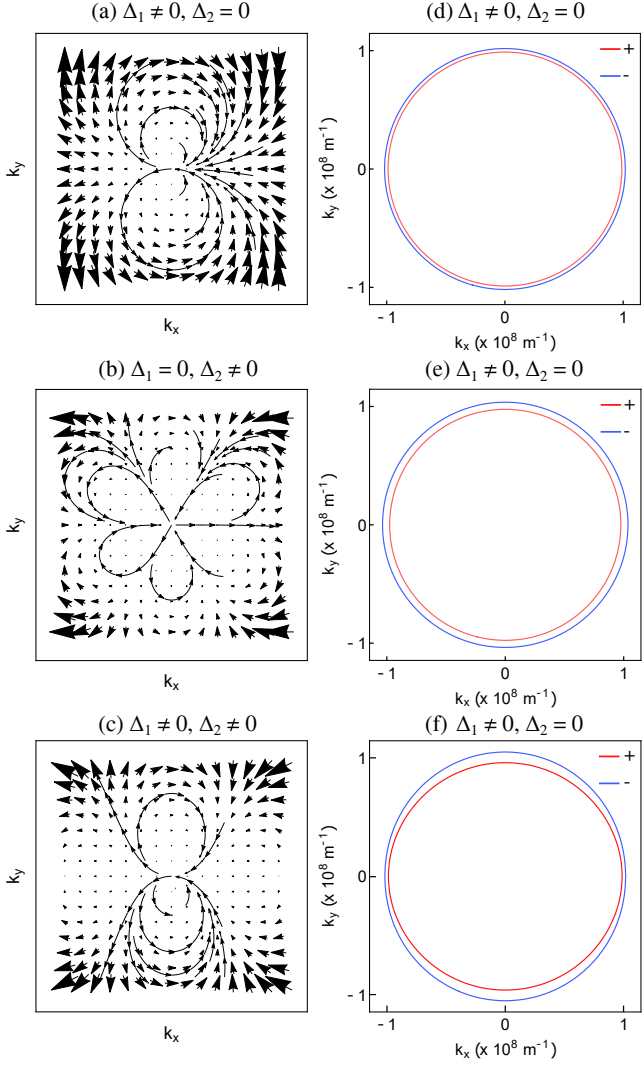}
	\caption{Schematics for the vector field $(\Omega_x,\Omega_y)$ with (a) $\Delta_{2} = 0$, (b) $\Delta_{1} = 0$, and (c) $\Delta_{1} k^2 \approx \Delta_{2} k^4$ with $B_{\parallel} = B_x$. The values for $\Delta_{1}$ and $\Delta_{2}$ are $\Delta_1 = 1 \times 10^{-17}$ meV m$^{2}$ T$^{-1}$, $\Delta_2 = -2 \times 10^{-33}$ meV m$^{4}$ T$^{-1}$, respectively. The Zeeman splitting for $p = 1.6 \times 10^{11}$ cm$^{-2}$ and  $B_{\parallel}$ = 1 T was found to be (d) $\sim 3\%$, (e) $\sim6\%$, and (f) $\sim6\%$ of the Fermi energy. Here, $+(-)$ refers to $m_J = +3/2$ $(-3/2)$ heavy holes subband.  The values for $\Delta_{1}$ and $\Delta_{2}$ are taken from Ref.~\cite{Miserev2017}.}
	\label{fig:Hamiltonian}
\end{figure}

Assuming that only the lowest (i.e. zero-node) heavy hole subbands are occupied, which is often the case experimentally, the band Hamiltonian with in-plane Zeeman terms is
\begin{equation}
	\label{eq:Zeeman_Hamiltonian}
	\begin{array}{rl}
		H_{0{\bm k}} &\dps= \frac{\hbar^2 k^2}{2m^* } + \Delta_{1} B_{+}k_+^2\sigma_- + \Delta_{2} B_{-}k_+^4\sigma_- + \mathrm{h.c.},
	\end{array}
\end{equation}
where the first term is the orbital term with $m^*$ an effective hole mass and the rest part, known as the Zeeman Hamiltonian, is equivalent to $\frac{1}{2}\boldsymbol{\sigma}\cdot\boldsymbol{\Omega_{\boldsymbol{k}}}$ having
\begin{equation}
	\hat{\bm \Omega}_{\bm k}  \dps = \frac{\Delta_{1} k^2 B_{\parallel} (c_{2\theta+\phi} + \Delta k^2 c_{4\theta -\phi}, s_{2\theta+\phi} + \Delta k^2 s_{4\theta -\phi} , 0) }{\sqrt{G(k)}},
\end{equation}
$k_{\pm} \equiv k_x \pm i k_y$, $k \equiv \sqrt{k_x^2 + k_y^2}$, and h.c. denotes hermitian conjugate for the Zeeman terms. The prefactors $\Delta_{1}$ and  $\Delta_{2}$ were calculated for specific structures in Ref.~\citenum{Miserev2017} and can be generically determined from ${\bm k}\cdot{\bm p}$ theory, the details of which we defer to a future publication. We also use $\Delta \equiv \Delta_2/\Delta_1$, while $B_{\pm} \equiv B_x \pm i B_y$, with $B_{x(y)}$ the $x(y)-$component of the magnetic field, $\sigma_{\pm} \equiv \sigma_{x} \pm i \sigma_y$, with $\sigma_{i}$ the Pauli spin matrices, $G(k) \equiv (\Delta_{1}^2 k^4 + \Delta_{2}^2 k^8 + 2 \Delta_{1} \Delta_{2} k^6 c_{2(\theta -\phi)})B_{\parallel}^2$, $\theta = \tan^{-1}(k_y/k_x)$, $\phi = \tan^{-1}(B_y/B_x)$, $c(s)$ refers the cosine(sine) trigonometric operators, and $\mathrm{h.c.}$ stands for Hermitian conjugate. We neglect the Dresselhaus spin-orbit interaction \cite{Dresselhaus1955} and cubic-symmetry Luttinger terms, since these do not change the interaction with the magnetic field. We consider cases where $B_{\parallel}$ is sufficiently small so that the lowest order Schrieffer-Wolff approximation for the Zeeman interaction (Eq.~\eqref{eq:Zeeman_Hamiltonian}) remains valid. At a typical density of $p = 1.6 \times 10^{11}$ cm$^{-2}$, Eq.~\eqref{eq:Zeeman_Hamiltonian} applies for $B_{\parallel} \leq 3.9$ T and $B_{\parallel} \leq 2.2$ T for GaAs and InAs, respectively. We ignore Landau levels since, for the range of $B_{\parallel}$ considered in the work, the magnetic length is at least one order of magnitude larger than the spatial overlap of the wave function, i.e. $\sqrt{\hbar/(e B_\parallel)} \gg \bkt{z}$. We do not include higher-order Zeeman interactions $\propto B_{\parallel}^3$ \cite{Roland} since they are approximately 3 orders of magnitude smaller than the $B$-linear term, meaning that the $B_{\parallel}^3$ term only becomes important when $B_{\parallel} \sim 30$ T.

The peculiar form of the Zeeman interaction in Eq.~\ref{eq:Zeeman_Hamiltonian} implies that the in-plane magnetic field in real space becomes an \textit{effective} magnetic field ${\bm \Omega}_{\bm k}$ in momentum space. We have plotted this effective field in Figs.~\ref{fig:Hamiltonian}a-c. As expected, the texture of the effective field and the energy dispersion are both strongly density-dependent. The $\Delta_{1}$ term has a winding number $w_{1} = 2$ and is dominant at lower densities whereas the $\Delta_{2}$ term has a winding number $w_{2} = 4$ and is dominant at higher densities. Figs.~\ref{fig:Hamiltonian}d-f show the Zeeman splitting of the heavy hole subbands, calculated using the dispersion relation $E(k) = \frac{\hbar^2 k^2}{2 m^*} \pm \sqrt{G(k)}$ for a typical experimental density of $p = 1.6 \times 10^{11}$ cm$^{-2}$ and $B_{\parallel} = B_x = 1$ T. If either $\Delta_{1}$ or $\Delta_{2}$ is zero (Fig.~\ref{eq:Zeeman_Hamiltonian}d-e) the energy dispersion is isotropic, while if the two are comparable in magnitude the dispersion is anisotropic (Fig.~\ref{eq:Zeeman_Hamiltonian}f). The anisotropy in the Fermi contours occurs because of the existence of two terms with different winding numbers. Finally we note that there exists a density and $\phi$ at which the term $\Delta_1 k^2$ exactly equals $\Delta_2 k^4$. At this point, the term $G(k)$ is zero at $\phi = \pi/2$ and $3\pi/2$, causing the perturbation expansion to fail. This case, which occurs at low densities, is not included in our calculation since it involves a laborious extension of the theory. From the physical standpoint, at the point where $G(k)$ becomes zero the two heavy hole subbands become degenerate, which contradicts our starting assumption that the two subbands are split by the Zeeman interaction and that they have different subband occupations as well as occupation probabilities. 

The momentum-dependent Zeeman terms produce sizable modifications of the longitudinal and Hall conductivities as $B_\parallel$ is increased, as well as a large anisotropy in both as the magnetic field is rotated in the plane of the sample. The strong anisotropy relative to the orientation of magnetic field is shown for GaAs and InAs in Fig.~\ref{fig:Results} which is the central result of this work. This quantum mechanical effect present for weak momentum scattering, is temperature-independent and has a strong density dependence. The angular dependence of the Hall coefficient $R_H$ can be used to identify the presence of the two Zeeman terms. Surprisingly, the Hall coefficient, a classical probe characterizing the interaction with an out-of-plane magnetic field, can be used to characterize the relativistic interaction with an \textit{in-plane} magnetic field. This naturally also in the presence of an in-plane magnetic field, $R_H$ due to an out of plane magnetic field is not simply dependent on the carrier density.

The physical reasons behind this effect are as follows. The in-plane magnetic field gives rise to a non-trivial spin texture and to a spin splitting, which renormalizes the Fermi velocity of the spin-split sub-bands, as well as inter- and intra-band scattering times. The net effect of these processes is a sizable renormalization of the Lorentz force acting on the holes. This is surprising for three fundamental reasons. Firstly\bw{,} $B_\parallel$ does not contribute directly to the Lorentz force but generates an effective momentum-dependent magnetic field (depending on the real magnetic field). Secondly, despite the presence of an effective magnetic field, the Berry curvature is zero here since the \textit{out-of-plane} Zeeman interaction is overwhelmed by disorder. Thirdly, quantum mechanical effects in charge transport, such as weak localization \cite{Datta1995} and the Altshuler-Aronov terms in the conductivity \cite{Altshuler1979, Altshuler1980}, are typically observed in diffusive samples only at very low temperatures. This is true even when the Fermi surface remains isotropic, which occurs if only one of the two Zeeman terms is present. We stress there is no Rashba spin-orbit coupling, while the Dresselhaus interaction will not alter the dependence of the results on the magnitude and orientation of $B_\parallel$.

\begin{figure}
	\centering
	\includegraphics[scale=0.8]{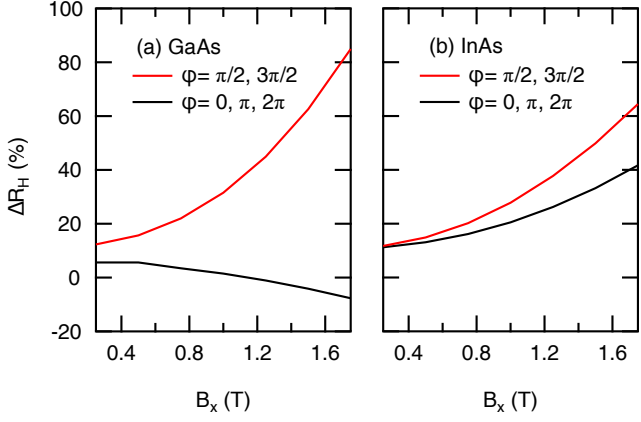}
	\caption{The extrema of $\Delta R_{\mathrm{H}}$ for (a) GaAs and (b) InAs as a function of the in-plane magnetic field $B_x$;  $k_F = 10^{8}$ m$^{-1}$.}
	\label{fig:RHall_extrema}
\end{figure}

Fig.~\ref{fig:RHall_extrema} shows the extrema of $\Delta R_H$ as a function of $B_x$. The correction $\Delta R_H$ of the Hall coefficient is maximal when $\phi = \pi/2, 3\pi/2$ and minimal when $\phi = 0, \pi, 2 \pi$. For GaAs, the quadratic-$k$ and quartic-$k$ Zeeman interactions are of the same order of magnitude. Hence, $\Delta R_H$ changes sign depending on $\phi$. In contrast, for InAs, the Zeeman interactions are dominated by the quartic-$k$ Zeeman term, hence its angular dependence of $\Delta R_H$ becomes less important than GaAs.

Experimentally, the quadratic-$k$ and quartic-$k$ Zeeman interactions can be distinguished in 1D holes by measuring the directional dependence of the electrical conductance \cite{Miserev2017a} and can be separately determined by magnetic focusing \cite{Bladwell2019}. On the other hand, the correction $\Delta R_H$ can be measured via the dependence of $R_H$ on $\phi$ in 2D holes, without requiring ultra-low temperatures, a large magnetic field, or optical setups. Our findings have consequences for quantum computation since electron dipole spin resonance needs both the spin-orbit interaction and a magnetic field for a spin-orbit qubit to be encoded as well as manipulated, and spin blockade depends sensitively on the in-plane magnetic field \cite{Jo-2017-PRB}. A strong $\Delta_2$ would mean that an in-plane magnetic field could be used to tune the Berry curvature and topological properties of two-dimensional holes proximity-coupled to a ferromagnet or an antiferromagnet \cite{Ohno-RMP-2014}. Furthermore, $B_{\parallel}$ could also be used to make one band nearly flat, increasing the effect of hole-hole interactions relevant for strongly correlated phases.

\begin{figure}
	\centering
	\includegraphics[scale=0.75]{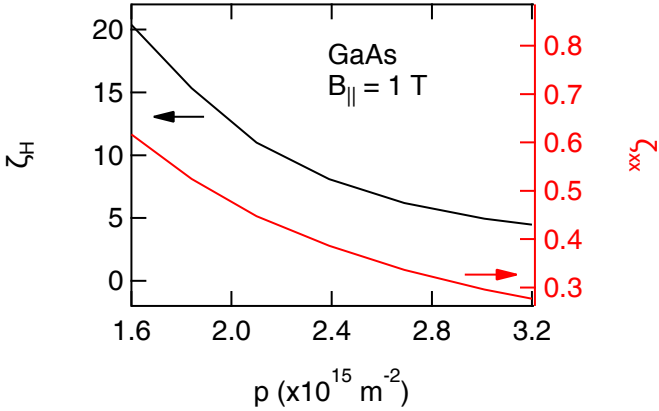}
	\caption{The anisotropy $\zeta_H$ and $\zeta_{xx}$ of the Hall resistance and longitudinal conductivity as a function of hole density, defined as $\zeta_H\equiv\frac{\Delta R_H(\pi/2)-R_H(0)}{R_H(0)}$ and $\zeta_{xx}\equiv\frac{\Delta \sigma_{xx}(\pi/2)-\sigma_{xx}(0)}{\sigma_{xx}(0)}$ respectively.} 
	\label{fig:Anisotropy_vs_dens}
\end{figure}

We briefly summarize the derivation steps of the magnetotransport coefficients, whose details can be found in the Supplemental Material. The density operator $\hat{\rho}$ obeys the quantum Liouville equation
\begin{equation} \label{QLE}
\frac{d \hat{\rho}}{dt}+\frac{i}{\hbar}[\hat{H},\hat{\rho}]=0.
\end{equation}
Projecting the density operator onto the states $\{\ket{\bm{k}s}\}$, where $\bm{k}$ is the wave vector and $s$ denotes the spin index, the matrix elements $\hat{\rho}_{{\bm k}{\bm k}'}\equiv \hat{\rho}^{ss'}_{{\bm k}{\bm k}'}=\langle {\bm k}s |\hat{\rho}|{\bm k}'s'\rangle$. Here we assume short-ranged uncorrelated impurities such that the average of potential over impurity configurations is $n_i \vert U_{\textbf{k}'\textbf{k}}\vert^2/V$, where $V$ is the crystal volume, $n_i$ is the impurity concentration, and $U_{\textbf{k}'\textbf{k}}$ is the matrix element of an impurity's potential. The central quantity in transport is the density matrix averaged over impurity configurations, which is also diagonal in wave vector, since the impurity average restores translational periodicity. We denote the impurity-averaged density matrix by $f_{\bm k}$. In the Born approximation for the impurity potential it satisfies the following quantum kinetic equation
\begin{equation}\label{Kinetic-e} 
\frac{df_{\bm k}}{dt}+\frac{i}{\hbar}\left[H_{0{\bm k}} + H_{\text{Z}}, f_{\bm k}\right]+\hat{J}(f_{\bm k})=\mathcal{D}_{E,{\bm k}}+\mathcal{D}_{L,{\bm k}},
\end{equation}
where $H_{\text{Z}}=g\mu_{\text{B}}\boldsymbol{\sigma}\cdot \textbf{B}$ having $g$, a material specific parameter and $\mu_{\text{B}}$, the Bohr magneton, $\hat{J}(f_{\bm k})$ is the scattering term in the Born approximation, given as 
\begin{equation} \label{J0}
\hat{J}(f_{\bm k})\!=\!\left\langle\frac{1}{\hbar^2}\int^{\infty}_0\!dt'[\hat{U},e^{-\frac{iH_0t'}{\hbar}}[\hat{U},\hat{f}(t)]e^{\frac{iH_0t'}{\hbar}}]\right\rangle_{{\bm k}{\bm k}},
\end{equation}
while $\mathcal{D}_{E,{\bm k}}=-(1/\hbar)e\textbf{E}\cdot \partial f_{\textbf{k}}/\partial\textbf{k}$ is the driving term due to an applied electric field $E$ and $\mathcal{D}_{L,{\bm k}}=(e/2\hbar)\{\hat{\textbf{v}}\times \textbf{B},\partial f_\textbf{k}/\partial\textbf{k}\}$ contains the effect of the Lorentz force due to the out-of-plane magnetic field.

The steady-state solution to Eq.~\eqref{Kinetic-e} can be found as follows. We perform a perturbation expansion up to first order in the electric and out-of-plane magnetic fields, and up to second order in the in-plane magnetic field, meaning that we keep terms up to order $B_{\parallel}^2$. Firstly, we decompose $f_{\bm k}$ as $f_{\bm k} = f_{0{\bm k}} + f_{E{\bm k}} + f_{L{\bm k}}$, where $f_{0{\bm k}}$ is the equilibrium density matrix, $f_{E{\bm k}}$ is a correction to first order in the electric field when $B_z = 0$, and $f_{L{\bm k}}$ is an additional correction first order in the electric and out-of-plane magnetic fields. Secondly, we use the equilibrium density matrix $f_{0{\bm k}}$ in Eq.~\eqref{Kinetic-e} with zero out-of-plane magnetic field, but finite in-plane magnetic field, to obtain $\mathcal{D}_{E,{\bm k}}$, from which we obtain $f_{E{\bm k}}$. Thirdly, using $f_{E{\bm k}}$ and with only $f_{L{\bm k}}$ on the right hand side of Eq.~\eqref{Kinetic-e}, we solve for $f_{L{\bm k}}$. Finally, the current density is found as the expectation value $ {\bm j}_{x, y}=e\text{Tr}\big[\hat{\bm v}_{x, y} f_{{\bm k}}\big]$, with $\hat{\bm v}= (1/\hbar) \, \partial H_{0{\bm k}}/\partial {\bm k}$, which yields the conductivity tensor $\sigma$. 

Following the above procedure, the longitudinal conductivity $\sigma_{xx}$ can be expressed as
\begin{widetext}
	\begin{equation}
	\sigma_{xx} = \frac{e^2\tau_p}{2\pi m^*}k^2_{\text{F}}\bigg[1- \left(\frac{m^{*2} B_{\parallel}^2}{\hbar^4}\right) \left(20 \Delta_{1}^2 + 104 k_F^4 \Delta_{2}^2 + 44 k_F^2 \Delta_{1} \Delta_{2} c_{2\phi}  + 2g(\phi) \right) \bigg]
	\end{equation}
where $k_F$ is the Fermi wave vector and for the case $B_{\parallel} = 0$, $k_F = \sqrt{2 \pi p}$, having $p$ is the 2DHG density, and
	\begin{equation}
	g(\phi) =  -\frac{2}{\pi}  \int_{0}^{2 \pi} d\theta c_{\theta} 
	\frac{I+J}{G(k_F)}\left(2 - \frac{k_F^2 B_{\parallel}^2 \Delta_{1}}{\sqrt{G(k_F)}}\right) \left(2\Delta_{1}^2 c_{\theta} + \Delta_{1}\Delta_{2} k_F^2 (2c_{3\theta - 2\phi} + 4c_{\theta-2\phi} ) + 4 \Delta_{2}^2k_F^4 c_{\theta}  \right),
	\end{equation}
\end{widetext}
with $I \equiv \Delta_2^2 k_F^8 B_\parallel^2$, and $J \equiv \Delta_1 \Delta_2 k_F^6 B_\parallel^2 c_{2(\theta-\phi)}$. The complete expressions for the transverse conductivity $\sigma_{xy}$ and Hall coefficient $R_{\mathrm{H}}$ are given in the Supplemental Material. The expressions for $\Delta \sigma_{xx}$, $\Delta \sigma_{xy}$ and $\Delta R_{\mathrm{H}}$ all contain terms $ \propto \cos[(\Delta w) \phi]$, where $\Delta w \equiv | {w_1 - w_2}| = 2$ is the difference in the winding number of the two Zeeman terms. Correspondingly, the period of the anisotropy in the magnetoresistance is $2 \pi / (\Delta w) = \pi$, which explains the angular dependence of the magnetoresistance in Figs.~\ref{fig:Results} and \ref{fig:RHall_extrema}.
 
As expected from the behavior of the dispersion at low and high densities (Fig.~\ref{fig:Hamiltonian}), the anisotropy in the magnetoresistance is strongly highly density dependent. At very high densities the anisotropy decreases, as seen in Fig.~\ref{fig:Anisotropy_vs_dens}, because the quartic Zeeman term $\propto \Delta_2$ overwhelms the quadratic, and the dispersion once again becomes approximately isotropic. Previous experiments probed the density dependence of the total Zeeman splitting in an out-of-plane magnetic field \cite{Marcellina2018}. This work shows that  the existence of the quartic term, with a winding number of 4,  can be detected by measuring the angular dependence of the in-plane magnetoresistance. 

To summarize, we have evaluated the longitudinal conductivity $\sigma_{xx}$ and low-field Hall coefficient $R_H$ of two-dimensional holes in GaAs and InAs symmetric quantum wells in an in-plane magnetic field, but with no Rashba spin-orbit interaction. In two-dimensional holes the strong spin-orbit interaction gives rise to a Zeeman interaction of the form $(\Delta_{1} B_{+}k_+^2\sigma_- + \Delta_{2} B_{-}k_+^4\sigma_- + \mathrm{h.c.})$, which causes the velocity operator, and hence $\sigma_{xx}$, $\sigma_{xy}$, and $R_H$, to depend strongly on the relative orientation of the electric and magnetic fields. This anisotropy in the magnetoresistance should be easily detected experimentally and should be observable in any materials with spin-orbit terms that have different winding numbers. 

Given the generality of our results, we expect that our work will stimulate further studies on the signatures of spin states in the classical transport properties of strongly spin-orbit coupled systems. Our work also has important implications for electron dipole spin resonance in the context of quantum computing, where the in-plane magnetic field acts as a tunable spin-orbit interaction through which the spin states can be manipulated. Finally, our work shows that one can probe the band topology from the magnetoresistance, with anisotropy period $2\pi/(\Delta w)$ where $\Delta w$ is the difference in the winding number of the spin-orbit terms.  Furthermore, one can determine from the magnetoresistance whether a type of spin-orbit interaction with a given winding number is dominant, or whether there are two spin-orbit terms of comparable magnitude that have a different winding number. In the latter case, the system is close to a point where a topological phase transition can occur.

This work was funded by the Australian Research Council Centre of Excellence in Future Low-Energy Electronics Technologies. We thank Hong Liu and Samuel Bladwell for helpful discussions.


%

\end{document}


\title{Supplement to ``Signatures of quantum mechanical Zeeman effect in classical transport due to topological properties of two-dimensional spin$-3/2$ holes"}

\author{E. Marcellina}
\altaffiliation{Present address: School of Physical and Mathematical Sciences, Nanyang Technological University, 21 Nanyang Link, Singapore 637371. Email: emarcellina@ntu.edu.sg}
\affiliation{School of Physics and Australian Research Council Centre of Excellence in Future Low-Energy Electronics Technologies, The University of New South Wales, Sydney 2052, Australia}

\author{Pankaj Bhalla}
\affiliation{Beijing Computational Science Research Center, Beijing 100193, China}

\author{A.~R.~Hamilton}
\affiliation{School of Physics and Australian Research Council Centre of Excellence in Future Low-Energy Electronics Technologies, The University of New South Wales, Sydney 2052, Australia}

\author{Dimitrie Culcer}
\affiliation{School of Physics and Australian Research Council Centre of Excellence in Future Low-Energy Electronics Technologies, The University of New South Wales, Sydney 2052, Australia}

\maketitle


\section{The effective spin-orbit field and velocity operator} 

The Zeeman Hamiltonian is given as

\begin{equation}   
H_{SO}= k^2  B_{\parallel} \big[\left( {\Delta_{1}}c_{2 \theta + \phi}+ k^2 {\Delta_{2}}c_{4 \theta -\phi} \right) \sigma_x + \left( {\Delta_{1}} s_{2 \theta + \phi} + k^2 {\Delta_{2}}s_{4 \theta -\phi}\right) \sigma_y \big] \equiv \frac{1}{2}{\bm \sigma}\cdot{\bm \Omega}_{\bm k}, 
 \end{equation}  
where $\Delta_{1}$ is the quadratic-$k$ prefactor, $\Delta_{2}$ is the quartic-$k$ Zeeman prefactor, $B_{\parallel} \equiv \sqrt{B_x^2 + B_y^2}$,  $B_{x(y)}$ is the $x(y)-$component of the magnetic field, $c(s)$ represents the trigonometric operator $\cos$($\sin$), $\sigma_{x,y}$ are the Pauli spin matrices, $\hat{\bm \Omega}_{\bm k}\cdot \hat{\bm \Theta}_{\bm k}=0$, $\theta = \text{arctan}(k_y/k_x)$, and  $\phi = \text{arctan}(B_y/B_x)$.  
 
The unit vectors corresponding to the effective spin-orbit field are:
\begin{equation}
  \ba
 \hat{\bm \Omega}_{\bm k}  &\dps =  k^2 B_{\parallel} (\Delta_{1} c_{2\theta+\phi} + \Delta_{2} k^2 c_{4\theta -\phi}, s_{2\theta+\phi} + \Delta_{1} k^2 s_{4\theta -\phi} , 0) / \sqrt{G(k)},\\[3ex]
 &\dps \equiv ( \Omega_x , \Omega_y  , 0)  \\[3ex]
 \hat{\bm \Theta}_{\bm k}  &\dps =  k^2 B_{\parallel}  [-\Delta_{1} s_{2\theta+\phi} - \Delta_{2} k^2 s_{4\theta -\phi}, \Delta_{1} c_{2\theta+\phi} + \Delta_{2} k^2 c_{4\theta -\phi} , 0] / \sqrt{G(k)} \\[3ex]
  &\dps \equiv ( \Theta_x , \Theta_y , 0 ) = ( - \Omega_y , \Omega_x , 0)  , 
 \ea
\end{equation} 
where
\begin{equation}
	\dps G(k) \equiv (\Delta_{1}^2 k^4 + \Delta_{2}^2 k^8 + 2 \Delta_{12} k^6 c_{2(\theta -\phi)})B_{\parallel}^2,
\end{equation}
and
\begin{equation}
	\Delta_{12} \equiv \Delta_{1} \Delta_{2}.
\end{equation}

The velocity operator contains contributions from the orbital motion and spin-orbit interaction, i.e.
\begin{equation}
	\ba
	\dps \hat{v}_x &= \frac{1}{\hbar} (\frac{\partial H_{\mathrm{orbital}}}{\partial k_x} + \frac{\partial H_{\mathrm{SO}}}{\partial k_x}) \\
	& = \frac{\hbar k_x}{m^*} \openone + \frac{B_{\parallel}}{\hbar}
	\left(
		\left[
				2 \Delta_{1} (k_x c_{\phi} - k_y s_{\phi}) + 4 \Delta_{2} (k_x^3 c_{\phi} - 3 k_x k_y^2 c_{\phi}  - k_y^3 s_{\phi} + 3 k_x^2 k_y s_{\phi})
		\right] \sigma_x\right.
		\\[3ex]
	&\left.		+ 	\left[	2 \Delta_{1} (k_y c_{\phi} + k_x s_{\phi}) + 4 \Delta_{2} (-k_y^3 c_{\phi} + 3 k_x^2 k_y c_{\phi}  - k_y^3 s_{\phi} + 3 k_x k_y^2 s_{\phi})	\right]  \sigma_y	\right)  \\[3ex]
	& =  \frac{\hbar k_x}{m^*} \openone + \frac{2 B_{\parallel} k}{\hbar} \left[ (\Delta_{1}c_{\theta+\phi} + 2 \Delta_{2} k^2 c_{3\theta - \phi} ) \sigma_{x} + (\Delta_{1}s_{\theta+\phi} + 2 \Delta_{2} k^2 s_{3\theta - \phi} ) \sigma_{y} \right].
	\ea
\end{equation}
Similarly, the $y-$component of the velocity operator is
\begin{equation}
	\dps \hat{v}_y =   \frac{\hbar k_y}{m^*} \openone + \frac{2 B_{\parallel} k}{\hbar} \left[ -(\Delta_{1}s_{\theta+\phi} + 2 \Delta_{2} k^2 s_{3\theta - \phi} ) \sigma_{x} + (\Delta_{1}c_{\theta+\phi} + 2 \Delta_{2} k^2 c_{3\theta - \phi} ) \sigma_{y} \right].
\end{equation}

{Note that the $x-$component of the spin-orbit velocity $\hat{v}_{\mathrm{SO},s} \equiv  \frac{\partial H_{\mathrm{SO}}}{\partial k_x}$  projected onto ${\bm \sigma}_{{\bm k}\parallel}$, where $\sigma_{{\bm k}\parallel} \equiv \Omega_x \sigma_{x} + \Omega_y \sigma_{y}$, reads:
\begin{equation}\label{eq:v_SO_x}
\hat{v}_{\mathrm{SO},x} = \frac{2 k^3 B_{\parallel}^2}{\hbar \sqrt{G(k)}}\left(\Delta_{1}^2 c_{\theta} + \Delta_{12} k^2 (2c_{\theta-2\phi} + c_{3\theta-2\phi}) + 2 \Delta_{2}^2 k^4 c_{\theta} \right) \sigma_{{\bm k}\parallel}.
\end{equation}

Similarly, the $y-$component of the spin-orbit velocity $\hat{v}_{\mathrm{SO},y} \equiv  \frac{\partial H_{\mathrm{SO}}}{\partial k_y}$ projected onto ${\bm \sigma}_{{\bm k}\parallel}$ reads:
\begin{equation}\label{eq:v_SO_y}
\hat{v}_{\mathrm{SO},y} = \frac{2 k^3 B_{\parallel}^2}{\hbar \sqrt{G(k)}}\left(\Delta_{1}^2 s_{\theta} + \Delta_{12} k^2 (-2s_{\theta-2\phi} + s_{3\theta-2\phi}) + 2 \Delta_{2}^2 k^4 s_{\theta} \right) \sigma_{{\bm k}\parallel}.
\end{equation}
We will need both $\hat{v}_{\mathrm{SO},x}$ and $\hat{v}_{\mathrm{SO},y} $ to evaluate the spin-orbit induced corrections to the longitudinal conductivity $\sigma_{xx}$ and Hall coefficient $R_H$, respectively.

\section{Scattering terms}

The scattering terms, in the first order Born approximation, are given as \cite{Culcer_InterbandCoh_PRB2017}
 \begin{equation}
 \label{eq:scattering_term_general}
 \ba 
 \hat{J}(f_{\bm k}) &\dps =\frac{n_i}{\hbar^2}\lim_{\eta\rightarrow 0}\int \frac{d^2{k'}}{(2\pi)^2}|\overline{U}_{{\bm k}{\bm k}'}|^2
\int^{\infty}_0 dt'e^{-\eta t'}\big\{e^{-iH_{{\bm k}'}t'/\hbar}(f_{\bm k}-f'_{\bm k})e^{iH_{\bm k}t'/\hbar}+\text{H.c.}\big\}\\[3ex]
&\dps =\frac{n_i}{\hbar^2}\lim_{\eta\rightarrow 0}\int \frac{d^2{k'}}{(2\pi)^2}|\overline{U}_{{\bm k}{\bm k}'}|^2 (n_{\bm k}-n_{{\bm k}'})
\int^{\infty}_0 dt'e^{-\eta t'}\big\{e^{-iH_{{\bm k}'}t'/\hbar}e^{iH_{\bm k}t'/\hbar}+e^{-iH_{{\bm k}}t'/\hbar}e^{iH_{{\bm k}'}t'/\hbar}\big\}\\[3ex]
&\dps +\frac{n_i}{2\hbar^2}\lim_{\eta\rightarrow 0}\int \frac{d^2{k'}}{(2\pi)^2}|\overline{U}_{{\bm k}{\bm k}'}|^2 ({\bm S}_{\bm k}-{\bm S}_{{\bm k}'})\cdot
\int^{\infty}_0 dt'e^{-\eta t'}\big\{e^{-iH_{{\bm k}'}t'/\hbar}{\bm \sigma}e^{iH_{\bm k}t'/\hbar}+e^{-iH_{{\bm k}}t'/\hbar}{\bm \sigma}e^{iH_{{\bm k}'}t'/\hbar}\big\},
\ea
 \end{equation}
where $n_i$ is the impurity density, $\overline{U}_{{\bm k}{\bm k}'}$ is the impurity potential, $n_{\bm k}$ and $S_{\bm k}$ are the scalar and spin components of the density matrix, i.e. $f_{\bm k} = n_{\bm k}\openone + S_{\bm k}$. With the following definitions:
\begin{equation}
\ba 
& C \equiv \frac{k^2 k'^2 B_{\parallel}}{\sqrt{G(k)H(k',\gamma)}}(\Delta_1^2 c_{2\gamma} + \Delta_2^2 k^2 k'^2  c_{4 \gamma} + \Delta_{12} (k^2 c_{2(\gamma-\theta+\phi)} + k'^2 c_{2(\theta -\phi + 2\gamma)})),\\[3ex]
& S \equiv  \frac{k^2 k'^2 B_{\parallel}}{\sqrt{G(k)H(k',\gamma)}}(\Delta_1^2 s_{2\gamma}  + \Delta_2^2 k^2 k'^2 s_{4\gamma}  + \Delta_{12} (k^2 s_{2(\gamma-\theta+\phi)} + k'^2 s_{2(\theta -\phi + 2\gamma)})),\\[3ex] 
\ea
\end{equation}
and
\begin{equation}
	H(k,\gamma)  \equiv (\Delta_{1}^2 k^4 + \Delta_{2}^2 k^8 + 2 \Delta_{12} k^6 c_{2(\theta +\gamma -\phi)}) B_{\parallel}^2,
\end{equation}
where $\gamma$ is the scattering angle, we evaluate the scattering terms (Eq.~\eqref{eq:scattering_term_general}) and obtain:
\begin{equation}
\ba
\hat{J}(n)  &\dps =\frac{\pi n_i}{2\hbar}\int\frac{d^2k'}{(2\pi)^2}|\overline{U}_{{\bm k}{\bm k}'}|^2(n_{\bm k}-n_{{\bm k}'})
(1 + C)\Big[2\delta(\epsilon_{0}-\epsilon'_{0})+\{\sqrt{G(k)}-\sqrt{H(k',\gamma)}\}^2\frac{\partial^{2} \delta(\epsilon_0-\epsilon'_0)}{\partial \epsilon^{2}_0}\Big]\\[3ex]
&\dps +\frac{\pi n_i}{2\hbar}\int\frac{d^2k'}{(2\pi)^2}|\overline{U}_{{\bm k}{\bm k}'}|^2(n_{\bm k}-n_{{\bm k}'})
(1 - C)\Big[2\delta(\epsilon_{0}-\epsilon'_{0})+\{\sqrt{G(k)}+\sqrt{H(k',\gamma)}\}^2\frac{\partial^{2} \delta(\epsilon_0-\epsilon'_0)}{\partial \epsilon^{2}_0}\Big],
\ea
\end{equation}
\begin{equation}
\ba 
\hat{J}(S)
&\dps =\frac{\pi n_i}{4\hbar}\int\frac{d^2k'}{(2\pi)^2}|\overline{U}_{{\bm k}{\bm k}'}|^2
\Big[(s_{{\bm k}\parallel}-s_{{\bm k}'\parallel})(1+C)\sigma_{{\bm k}\parallel}+(s_{{\bm k}\parallel}-s_{{\bm k}'\parallel})S\sigma_{{\bm k}\perp}+(s_{{\bm k}\perp}+s_{{\bm k}'\perp})\sigma_{{\bm k}\parallel}S\\[3ex]
&\dps +(s_{{\bm k}\perp}+s_{{\bm k}'\perp})(1-C)\sigma_{{\bm k}\perp}\Big]\Big[2\delta(\epsilon_{0}-\epsilon'_{0})+\{\sqrt{G(k)}+\sqrt{H(k',\gamma)}\}^2\frac{\partial^{2} \delta(\epsilon_0-\epsilon'_0)}{\partial \epsilon^{2}_0}\Big] \\[3ex]
&\dps +\frac{\pi n_i}{4\hbar}\int\frac{d^2k'}{(2\pi)^2}|\overline{U}_{{\bm k}{\bm k}'}|^2
\Big[(s_{{\bm k}\parallel}+s_{{\bm k}'\parallel})(1-C)\sigma_{{\bm k}\parallel}-(s_{{\bm k}\parallel}+s_{{\bm k}'\parallel})S\sigma_{{\bm k}\perp}-(s_{{\bm k}\perp}-s_{{\bm k}'\perp})\sigma_{{\bm k}\parallel}S\\[3ex]
&\dps +(s_{{\bm k}\perp}-s_{{\bm k}'\perp})(1{+}C)\sigma_{{\bm k}\perp}\Big]\Big[2\delta(\epsilon_{0}-\epsilon'_{0})+\{\sqrt{G(k)}+\sqrt{H(k',\gamma)}\}^2\frac{\partial^{2} \delta(\epsilon_0-\epsilon'_0)}{\partial \epsilon^{2}_0}\Big],
\ea
\end{equation}
\begin{equation}
\ba
\hat{J}_{S\rightarrow n}(S) &\dps =\frac{\pi n_i}{4\hbar}\int\frac{d^2k'}{(2\pi)^2}|\overline{U}_{{\bm k}{\bm k}'}|^2
\Big[(s_{{\bm k}\parallel}-s_{{\bm k}'\parallel})(1+C)+(s_{{\bm k}\perp}+s_{{\bm k}'\perp})S\gamma\Big]  
\Big[2(\sqrt{G(k)}-\sqrt{H(k',\gamma)})\pd{}{\epsilon_0}\delta(\epsilon_{0}-\epsilon'_{0})\Big]\\[3ex]
&\dps {+\frac{\pi n_i}{4\hbar}\int\frac{d^2k'}{(2\pi)^2}|\overline{U}_{{\bm k}{\bm k}'}|^2
	\Big[(s_{{\bm k}\parallel}+s_{{\bm k}'\parallel})(1-C)-(s_{{\bm k}\perp}-s_{{\bm k}'\perp})S\gamma\Big]  
	\Big[2(\sqrt{G(k)}+\sqrt{H(k',\gamma)})\pd{}{\epsilon_0}\delta(\epsilon_{0}-\epsilon'_{0})\Big]},
\ea
\end{equation}
\begin{equation}
\ba 
\hat{J}_{n\rightarrow S}(n) 
&\dps =\frac{\pi n_i}{2\hbar}\int\frac{d^2k'}{(2\pi)^2}|\overline{U}_{{\bm k}{\bm k}'}|^2(n_{\bm k}-n_{{\bm k}'})\Big[\sigma_{{\bm k}\parallel}(1+C)+\sigma_{{\bm k} \perp}S\Big]\Big[2(\sqrt{G(k)}-\sqrt{H(k',\gamma)})\pd{}{\epsilon_0}\delta(\epsilon_{0}-\epsilon'_{0})\Big] \\[3ex] 
&\dps +\frac{\pi n_i}{2\hbar}\int\frac{d^2k'}{(2\pi)^2}|\overline{U}_{{\bm k}{\bm k}'}|^2(n_{\bm k}-n_{{\bm k}'})\Big[\sigma_{{\bm k}\parallel}(1-C)-\sigma_{{\bm k} \perp}S\Big]  \Big[2(\sqrt{G(k)}+\sqrt{H(k',\gamma)})\pd{}{\epsilon_0}\delta(\epsilon_{0}-\epsilon'_{0})\Big].
\ea
\end{equation} 

Throughout this work, we assume short-range impurities, so that the scattering potential $\overline{U}_{{\bm k}{\bm k}'} \equiv U$ is constant and $\hat{J}_{S \rightarrow n}$ is zero, so that the momentum relaxation time $\tau_p$ and spin relaxation time $\tau_s$ are given as follows:
\begin{equation}
\ba
\dps	\tau_p &= \frac{\hbar^3}{m^* n_i U}, \\
\dps	\tau_s &= 2\frac{\hbar^3}{m^* n_i U} = 2 \tau_p. \\
\ea
\end{equation}
  
\section{Solving for the longitudinal conductivity}

\subsection{$\mathcal{D}^{(0)}$}
For the zeroth order driving terms we have: 
\begin{equation}
\ba
&\dps \frac{d n^{(0)}_{\bm k}}{dt}+\hat{J}^{(0)}_{n\rightarrow n} (n^{(0)}_{\bm k})=\mathcal{D}^{(0)}_{{\bm k}n}, \quad \hat{J}^{(0)}_{n\rightarrow n}(n^{(0)}_{\bm k}) =\frac{\pi n_i}{2\hbar}\int\frac{d^2k'}{(2\pi)^2}|\overline{U}_{{\bm k}{\bm k}'}|^2(n^{(0)}_{\bm k}-n^{(0)}_{{\bm k}'}) 4\delta(\epsilon_0-\epsilon'_0).\\[3ex]
&\dps \frac{d S^{(0)}_{{\bm k}\parallel}}{dt}+P_{\parallel}\hat{J}^{(0)}_{S\rightarrow S} (S^{(0)}_{{\bm k}\parallel})
=0,\\[3ex]
&\dps \frac{d S_{{\bm k}\perp}}{dt}+\frac{i}{\hbar}\big[H_{0{\bm k}},S_{{\bm k}\perp}\big]=\mathcal{D}_{{\bm k}\perp}.
   \ea
\end{equation}
\begin{equation}
	\label{eq:zeroth_order}
  \ba
 &\dps  n^{(0)}_{\bm k}=\frac{e{\bm E}\cdot \hat{\bm k}}{\hbar}\tau_p\Big[\frac{\hbar^2k}{m^*}\delta(\epsilon_0-\epsilon_{\text{F}})\Big], \\[3ex]
 &\dps \quad S^{(0)}_{{\bm k}\parallel}=0. \\[3ex]
\ea
 \end{equation}
The expressions in Eq.~\eqref{eq:zeroth_order} are consistent with the expectation that in the zeroth order spin does not contribute to conduction. 


 \subsection{$\mathcal{D}^{(1)}$}
  
For the first order term, there are two contributions to the  $S^{(1)}_{{\bm k}\parallel}$, namely one from the electric field  driving term, and another from the coupled scattering term $P_{\parallel}\hat{J}_{n \rightarrow S} (n^{(0)}_{\bm k})$. More explicitly,
\begin{equation}
  \ba
  &\dps \frac{d n^{(1)}_{\bm k}}{dt}+\hat{J}_{n\rightarrow n} (n^{(1)}_{\bm k}) =-\hat{J}_{S \rightarrow n} (S^{(0)}_{{\bm k}\perp})=0,\\[3ex]
  &\dps \frac{d S^{(1)}_{{\bm k}\parallel}}{dt}+P_{\parallel}\hat{J}_{S\rightarrow S} (S^{(1)}_{{\bm k}\parallel})+P_{\parallel}\hat{J}_{S\rightarrow S}(S^{(0)}_{{\bm k}\perp})+P_{\parallel}\hat{J}_{n \rightarrow S} (n^{(0)}_{\bm k})
=\mathcal{D}^{(1)}_{{\bm k}}. \\[3ex]
\ea
\end{equation} 

\begin{equation}
  \ba
\dps  P_{\parallel}\hat{J}_{n \rightarrow S} (n^{(0)}_{\bm k}) &\dps =\frac{\pi n_i}{2\hbar}\int\frac{d^2k'}{(2\pi)^2}|\overline{U}_{{\bm k}{\bm k}'}|^2(n^{(0)}_{\bm k}-n^{(0)}_{{\bm k}'})\big[4  \sqrt{G(k)}\pd{\delta(\epsilon_0-\epsilon'_0)}{\epsilon_0}-4  \sqrt{H(k',\gamma)}{C}\pd{\delta(\epsilon_0-\epsilon'_0)}{\epsilon_0}\big]\sigma_{{\bm k}\parallel}\\[3ex]
&\dps  =\frac{\pi n_i}{2\hbar}\int\frac{k'dk'd\theta'}{(2\pi)^2}|\overline{U}_{{\bm k}{\bm k}'}|^2n^{(0)}_k(c_{\theta}-c_{\theta'})\big\{4 \sqrt{G(k)}\frac{m^{*2}}{\hbar^4k}\Big[\frac{1}{k}\pd{\delta(k-k')}{k}-\frac{\delta(k-k')}{k^2}\Big]\big\}\sigma_{{\bm k}\parallel}\\[3ex]
&\dps -\frac{\pi n_i}{2\hbar}\int\frac{k'dk'd\theta'}{(2\pi)^2}|\overline{U}_{{\bm k}{\bm k}'}|^2n^{(0)}_k(c_{\theta}-c_{\theta'})\big\{4\sqrt{H(k',\gamma)} {C}\frac{m^{*2}}{\hbar^4k}\Big[\frac{1}{k}\pd{\delta(k-k')}{k}-\frac{\delta(k-k')}{k^2}\Big]\big\}\sigma_{{\bm k}\parallel}\\[3ex]
 &\dps = 0.
\ea
\end{equation}
\begin{equation}
\ba
&\dps  P_{\parallel}\hat{J}_{S\rightarrow S}(S^{(1)}_{{\bm k}\parallel})=\frac{\pi n_i}{4\hbar}\int\frac{d^2k'}{(2\pi)^2}|\overline{U}_{{\bm k}{\bm k}'}|^2(s^{(1)}_{{\bm k}\parallel}-s^{(1)}_{{\bm k}'\parallel}{C}) 4\delta(\epsilon_0-\epsilon'_0)\sigma_{{\bm k}\parallel}=\mathcal{D}^{(1)}_{\bm k}-P_{\parallel}\hat{J}_{n \rightarrow S} (n^{(0)}_{\bm k})\\[3ex]
&\equiv s_{\bm k_\parallel}^{(1)}\sigma_{{\bm k}\parallel} \frac{1}{\tau_s}
\ea
\end{equation}

\subsection{$\mathcal{D}^{(2)}$}

For the second order term $n^{(2)}_{\bm k}$, the effective driving term  $D^{(2),\mathrm{eff}}_{\bm k}$ has  three contributions: one is from the driving term, another from the coupled scattering term $\hat{J}_{S \rightarrow n}(S^{(1)}_{\bm k})$, and the last one is $\hat{J}^{(2)}_{n\rightarrow n}(n^{(0)}_{\bm k})$.

\begin{equation}
\ba
&\dps \frac{d n^{(2)}_{\bm k}}{dt}+\hat{J}^{(0)}_{n\rightarrow n} (n^{(2)}_{\bm k}) =\mathcal{D}^{(2)}_{\bm k}(B_{\parallel}^2 )-\hat{J}_{S \rightarrow n} (S^{(1)}_{{\bm k}\parallel})-\hat{J}^{(2)}_{n\rightarrow n} (n^{(0)}_{\bm k})= D^{(2),\mathrm{eff}}_{\bm k}, 
\ea
\end{equation}  
with
\begin{equation}
\ba
\dps \hat{J}_{S\rightarrow n}(S^{(1)}_{\bm k}) &\dps =\frac{\pi n_i}{4\hbar}\int\frac{d^2k'}{(2\pi)^2}|\overline{U}_{{\bm k}{\bm k}'}|^2(s^{(1)}_{{\bm k}\parallel}-s^{(1)}_{{\bm k}'\parallel})(1+C)2(\sqrt{G(k)}-\sqrt{G(k')})\pd{\delta(\epsilon_0-\epsilon'_0)}{\epsilon_0} \\[3ex]
&\dps +\frac{\pi n_i}{4\hbar}\int\frac{d^2k'}{(2\pi)^2}|\overline{U}_{{\bm k}{\bm k}'}|^2
  \Big[(s^{(1)}_{{\bm k}\parallel}+s^{(1)}_{{\bm k}'\parallel})(1-C)\Big]2(\sqrt{G(k)} + \sqrt{G(k')})\pd{\delta(\epsilon_{0}-\epsilon'_{0})}{\epsilon_0}\\[3ex]
&\dps =\frac{\pi n_i}{4\hbar}\int\frac{d^2k'}{(2\pi)^2}|\overline{U}_{{\bm k}{\bm k}'}|^2(1+C)s^{(1)}_{{k}\parallel}(c_{\theta}-c_{\theta'})2(\sqrt{G(k)}-\sqrt{G(k')})\frac{m^{*2}}{\hbar^4k}\Big[\frac{1}{k}\pd{\delta(k-k')}{k}-\frac{\delta(k-k')}{k^2}\Big] \\[3ex]
&\dps {+\frac{\pi n_i}{4\hbar}\int\frac{d^2k'}{(2\pi)^2}|\overline{U}_{{\bm k}{\bm k}'}|^2(1-C)
  s^{(1)}_{k\parallel}\Big[(c_{\theta}+c_{\theta'})\Big]  
  2(\sqrt{G(k)}+\sqrt{G(k')})\frac{m^{*2}}{\hbar^4k}\Big[\frac{1}{k}\pd{\delta(k-k')}{k}-\frac{\delta(k-k')}{k^2}\Big]}\\[3ex]
&\dps =  0,
 \ea  
\end{equation}
and
\begin{equation}
\ba
  \dps  \hat{J}^{(2)}(n^{(0)}_{\bm k}) &\dps=\frac{\pi n_i}{2\hbar}\int\frac{d^2k'}{(2\pi)^2}|\overline{U}_{{\bm k}{\bm k}'}|^2(n^{(0)}_{\bm k}-n^{(0)}_{{\bm k}'})
  \Big[{2 G(k)+2  H(k',\gamma)}-4 C \sqrt{G(k)H(k',\gamma)}\Big]\frac{\partial^{2} \delta(\epsilon_0-\epsilon'_0)}{\partial \epsilon^{2}_0}\\[3ex]
  &\dps  =\frac{\pi n_i}{2\hbar}\int\frac{d^2k'}{(2\pi)^2}|\overline{U}_{{\bm k}{\bm k}'}|^2 n^{(0)}_k(c_{\theta}-c_{\theta'})
\\[3ex]
&\dps  \times \Big\{  \big[  {2  G(k)+2 H(k',\gamma)}-4 C \sqrt{G(k)H(k',\gamma)} \big] \frac{m^{*3}}{\hbar^6}\Big[\frac{1}{k^3}\frac{\partial^{2}\delta(k-k')}{\partial^{2}k}-\frac{3}{k^4}\pd{\delta(k-k')}{k}+\frac{3\delta(k-k')}{k^5}\Big]
\Big\}\\[3ex]
&\dps  = n^{(0)}_{\bm k}\frac{n_im^{*3}}{2\pi\hbar^7} B_{\parallel}^2 \Big[(4 \Delta_{1}^2 + 24 k^4 \Delta_{2}^2)\xi(\gamma) + 4 k^2 \Delta_{12} \Gamma(\gamma) c_{2(\theta-\phi)} -20 k^4 \Delta_{2}^2  \Gamma(\gamma) \Big].
\\[3ex]
 \ea  
\end{equation}


We then solve the kinetic equation for the charge part
\begin{equation}
\hat{J}^{(2)}(n^{(2)}_{\bm k})=\frac{\pi n_i}{2\hbar}\int\frac{d^2k'}{(2\pi)^2}|\overline{U}_{{\bm k}{\bm k}'}|^2(n^{(2)}_{\bm k}-n^{(2)}_{{\bm k}'})
\Big[4\delta(\epsilon_{0}-\epsilon'_{0})\Big]= D^{(2),\mathrm{eff}}_{\bm k}.
\end{equation} 
 
\begin{equation}
\ba
&\dps n^{(2)}_{{\bm k}}= \tau_p\Bigg\{
 \frac{e{\bm E}\cdot\hat{\bm k}}{2\hbar}\Big[ \frac{4}{k}  \pd{\delta(\epsilon_0-\epsilon_{\text{F}})}{\epsilon_0} \left(G(k) + {\Delta_{2}} B_{\parallel}^2k^6(k^2 \Delta_{2} + \Delta_{1} c_{2(\theta-\phi)})\right) \Big]\\[3ex]
&\dps - \frac{e{\bm E}\cdot\hat{\bm \theta}}{2\hbar}\Big(\frac{4}{k}  \pd{\delta(\epsilon_0-\epsilon_{\text{F}})}{\epsilon_0} \Delta_{12} B_{\parallel}^2 k^6  s_{2(\theta-\phi)} \Big) - n^{(0)}_{\bm k}\frac{n_im^{*3}}{2\pi\hbar^7}[(4 \Delta_{1}^2 + 24 k^4 \Delta_{2}^2)\xi(\gamma)\Big] \Bigg\}.
\ea
\end{equation} 

\subsection{Expression for the longitudinal conductivity}

The current density operator is given as:
\begin{equation}
	\label{eq:current_density_operator}
	{\bm j}=e \text{Tr}\Big[{\bm \hat{v}} . \Big\{(n^{(0)}_{\bm k}+n^{(2)}_{\bm k})\openone+ s^{(1)}_{{\bm k}\parallel}\sigma_{{\bm k}\parallel} +  s^{(1)}_{{\bm k}\perp}\sigma_{{\bm k}\perp} \Big\}\Big].
\end{equation} 

The longitudinal conductivity, evaluated using Eq.~\eqref{eq:current_density_operator}, is thus given by
\begin{equation}
	\label{eq:sigma_xx}
	\sigma_{xx} = \frac{e^2\tau_p}{2\pi m^*}k^2_{\text{F}}\Big[1- \left(\frac{m^{*2} B_{\parallel}^2}{\hbar^4}\right) \left(20 \Delta_{1}^2 + 104 k_F^4 \Delta_{2}^2 + 44 k_F^2 \Delta_{12} c_{2\phi}  + 2g(\phi) \right) \Big],
\end{equation}
where
\begin{equation}
g(\phi) \equiv  -\frac{2}{\pi}  \int_{0}^{2 \pi} d\theta c_{\theta} 
\frac{I+J}{G(k_F)}\left(2 - \frac{k_F^2 B_{\parallel}^2 \Delta_{1}}{\sqrt{G(k_F)}}\right) \left(2\Delta_{1}^2 c_{\theta} + \Delta_{12} k_F^2 (2c_{3\theta} + 4c_{\theta-2\phi} ) + 4 \Delta_{2}^2k_F^4 c_{\theta}  \right).
\end{equation}.

\section{Solving for the Hall coefficient}


The Lorentz driving term is
\begin{equation}
\mathcal{D}_L=\frac{1}{2}\frac{e}{\hbar }\Big\{\hat{\bm v}\times {\bm B}, \pd{\rho_{E{\bm k}}}{{\bm k}}\Big\}.
\end{equation}

We switch coordinate from  $n(k_x,k_y)$ to $n(r,\theta)$ with $\pd{n}{k_x}=\pd{n}{k}c_{\theta}-\pd{n}{\theta}\frac{s_{\theta}}{k};\pd{n}{k_{y}}=\pd{n}{k}s_{\theta}+\pd{n}{\theta}\frac{c_{\theta}}{k}$, and obtain 

\begin{equation}
\mathcal{D}_{L,n} = 
\frac{e B_z}{m^*}(n^{(0)}_k+n^{(2)}_k)s_{\theta}+\frac{eB_z}{\hbar}\left[v_{\mathrm{SO,y}} \left(\pd{s^{(1)}_{k,\parallel}}{k}c_{\theta} - \pd{s^{(1)}_{k,\parallel}}{\theta} \frac{s_{\theta}}{k} \right) - v_{\mathrm{SO,x}} \left(\pd{s^{(1)}_{k,\parallel}}{k}s_{\theta} + \pd{s^{(1)}_{k,\parallel}}{\theta} \frac{c_{\theta}}{k} \right)\right],
\end{equation}
and 
\begin{equation}
\mathcal{D}_{L,S} =
\left( \frac{e B_z}{m^*}\left(-\frac{\partial s^{(1)}_{k,\parallel}}{\partial \theta}\right)+\frac{eB_z}{\hbar}\left[v_{\mathrm{SO,y}} \left(\pd{n^{(0)}_{k,\parallel}}{k}c_{\theta} - \pd{n^{(0)}_{k,\parallel}}{\theta} \frac{s_{\theta}}{k} \right) - v_{\mathrm{SO,x}} \left(\pd{n^{(0)}_{k,\parallel}}{k}s_{\theta} + \pd{n^{(0)}_{k,\parallel}}{\theta} \frac{c_{\theta}}{k} \right)\right] \right) \sigma_{{\bm k}\parallel}.
\end{equation}

{Using the kinetic equations} 
\begin{equation}\label{eq:kinetic_Lorentz_driving}
\ba
&\dps \frac{d n_{\bm k}}{dt}+\hat{J}_{n\rightarrow n} (n_{\bm k}) =\mathcal{D}_{L,n},\\[3ex]
&\dps \frac{d S_{{\bm k}\parallel}}{dt}+P_{\parallel}\hat{J}_{S \rightarrow S} (S_{{\bm k}\parallel})
=\mathcal{D}_{L,{\bm k}\parallel},\\[3ex]
\ea
\end{equation} 
{the steady-state solution to Eq. \eqref{eq:kinetic_Lorentz_driving} reads}
\begin{equation}
\label{eq:steady_state_Hall}
\ba
&\dps {n}_{L,n_{\bm k}}=\mathcal{D}_{L,n} \tau_p,\\[3ex]
&\dps {S}_{L,S_{\parallel,{\bm k}}}=\mathcal{D}_{L,S} \tau_s.\\[3ex]
\ea
\end{equation}

\subsection{Transverse conductivity}

Using Eqs.~\eqref{eq:current_density_operator}, ~\eqref{eq:kinetic_Lorentz_driving}, and ~\eqref{eq:kinetic_Lorentz_driving}, we evaluate the transverse conductivity. We define the following:
\begin{equation}
\ba
G &\equiv G(k_F) \\
H &\equiv H(k_F) \\
I &\equiv \Delta_2^2 k_F^8 B_{\parallel}^2 \\
J &\equiv \Delta_{12} k_F^6 B_{\parallel}^2 c_{2(\theta-\phi)} \\
v_x &\equiv  \Delta_1^2 c_{\theta + \gamma} + \Delta_{12} k_F^2 (c_{3 \theta + \gamma-2 \phi}+2 c_{\theta + \gamma-2 \phi})+2 \Delta_2^2 k_F^4 c_{\theta + \gamma}\\
v_y &\equiv  \Delta_1^2 s_{\theta + \gamma} - \Delta_{12} k_F^2 s_{(3 \theta + \gamma-2 \phi}+2  s_{\theta + \gamma-2 \phi})+2 \Delta_2^2 k_F^4 s_{\theta + \gamma},\\
\ea
\end{equation}

\begin{center}
	\begin{equation}
	\ba
	F_1 &\equiv \left[20-\frac{36 I+32 J}{G}+10 \frac{(I+J) (\Delta_1 B_{\parallel} k_F^2)}{\sqrt{G}^3}-(\frac{I+J}{G})^2 \left(6-\frac{4 \Delta_1 B_{\parallel} k_F^2} {\sqrt{G}}\right)\right] s_{2 (\theta + \gamma-\phi)} s_{2 (\theta + \gamma)} \\
	F_2 &\equiv -2 \frac{k_F^6 \Delta_{12}}{G} \left[ \left( 2 \frac{\Delta_1 B_{\parallel}^2 k_F^2}{\sqrt{G}}-4 \frac{I+J}{G} \frac{\Delta_1 B_{\parallel} k_F^2 }{\sqrt{G}} \right) \left(\Delta_1^2+\Delta_{12} k_F^2 c_{2 (\theta + \gamma-\phi)} +  2 \Delta_2^2 k_F^4 +  2 \Delta_{12} k_F^2 c_{2 (\theta + \gamma-\phi)}+8 \Delta_2^2 k_F^4\right) \right] \times \\
	& s_{2 (\theta + \gamma-\phi)} c_{\theta + \gamma} s_{\theta + \gamma} \\
	F_3 &\equiv -s^2_{\theta + \gamma} ((2 \Delta_1^2+4 \Delta_{12} k_F^2 c_{2 (\theta + \gamma-\phi)}+12 \Delta_2^2 k_F^4)-2 \frac{I+J}{G}  (2-\frac{\Delta_1 k_F^2 B_{\parallel}}{\sqrt{G}}) \left(\Delta_1^2+\Delta_{12} k_F^2 c_{2 (\theta + \gamma-\phi)} + 2 \Delta_2^2 k_F^4 \right)) \\
	F_4 &\equiv 2 \frac{k_F^6 \Delta_{12}}{G} s_{2 (\theta + \gamma-\phi)} c_{\theta + \gamma} \left[\left(-2 \frac{\Delta_1 k_F^2 B_{\parallel}}{\sqrt{G}}+4 \frac{I+J}{G}\frac{\Delta_1 k_F^2 B_{\parallel}}{\sqrt{G}} \right) v_y + 4 \Delta_{12} k_F^2 s_{\theta + \gamma-2 \phi} + 2 \Delta_{12} k_F^2 s_{3 \theta + \gamma-2 \phi}+8 \Delta_2^2 k_F^4 s_{\theta + \gamma} \right] \\
	F_5 &\equiv -s_{\theta + \gamma} \left[2 \Delta_1^2 s_{\theta + \gamma}-8 \Delta_{12} k_F^2 s_{\theta + \gamma-2 \phi}+4 \Delta_{12} k_F^2 s_{3 \theta + \gamma-2 \phi}+12 \Delta_2^2 k_F^4 s_{\theta + \gamma}+2 \frac{I+J}{G} \left(2-\frac{\Delta_1 B_{\parallel} k_F^2}{\sqrt{G}} \right) v_y\right] \\
	F_6 &\equiv \frac{k_F^4}{G} v_y^2 s_{\theta + \gamma}^2+\left(\frac{4}{G}+4 \frac{I+J}{G^2}\right) k_F^4 c_{\theta + \gamma}^2 v_y^2-7 \frac{k_F^4}{G} v_y^2 c_{\theta + \gamma}^2-2 (v_y \left(2 \Delta_{12} k_F^6 (-2 s_{\theta + \gamma-2 \phi} + s_{3 \theta + \gamma-2 \phi}) +  8 \frac{\Delta_2^2 k_F^8 s_{\theta + \gamma}}{G} \right) c_{\theta + \gamma}^2) \\
	F_7 &\equiv \frac{k_F^4}{G} v_y v_x c_{\theta + \gamma} s_{\theta + \gamma}+\left(\frac{4}{G}+4 \frac{I+J}{G^2}\right) k_F^4 c_{\theta + \gamma} s_{\theta + \gamma} v_y v_x -7 \frac{k_F^4}{G} v_y v_x c_{\theta + \gamma} s_{\theta + \gamma} +  \frac{1}{G}\left[v_{x} (2 \Delta_{12} k_F^6 (-2 s_{\theta + \gamma-2 \phi}+s_{3 \theta + \gamma-2 \phi}) + \right. \\
	& + \left. 8 \Delta_2^2 k_F^8 s_{\theta + \gamma})- (v_y (2 \Delta_{12} k_F^6 (2 c_{\theta + \gamma-2 \phi}+c_{3 \theta + \gamma-2 \phi})+8 \Delta_2^2 k_F^8 c_{\theta + \gamma})) c_{\theta + \gamma} s_{\theta + \gamma}\right],
	\ea
	\end{equation}	
\end{center}

\begin{equation}
\ba
L_1 &\equiv -(12 \Delta_1^2 + 56 \Delta_2^2 k_F^4 + 18 \Delta_{12} k_F^2  c_{2 \phi}) \\
L_2 &\equiv (4/\pi) \Delta_{12} k_F^2 \int_{0}^{2\pi}F_1 d \gamma \\
L_3 &\equiv (4/\pi) \int_{0}^{2\pi}(F_2+F_3) d \gamma\\  
L_4 &\equiv  (8/\pi) \int_{0}^{2\pi}(F_4+F_5) d \gamma\\
L_5 &\equiv  (8/\pi) \int_{0}^{2\pi}F_6 d \gamma\\
L_6 &\equiv  (8/\pi) \int_{0}^{2\pi}F_7 d \gamma\\
\ea
\end{equation}	
which finally gives
\begin{equation}
\label{eq:sigma_xy}
\sigma_{xy} = L_1 + L_2 + L_3 + L_4 + L_5 + L_6.
\end{equation}
\subsection{Hall coefficient}

The Hall coefficient $R_H$ at low magnetic fields is obtained by using 
\begin{equation}
	R_H = \lim_{B_z \rightarrow 0} \frac{\sigma_{xy}}{B_z(\sigma_{xx}^2 + \sigma_{xy}^2)},
\end{equation}
with $\sigma_{xx}$ given in Eq.~\eqref{eq:sigma_xx} and $\sigma_{xy}$ in Eq.~\eqref{eq:sigma_xy}. Note that in the absence of spin-orbit interaction, the Hall coefficient is $R_H =\frac{1}{pe}$ with $p$ hole density, $\sigma_{xx}=\frac{e^2\tau_p}{2\pi m^*}k^2_{\text{F}}$. This can be verified by 
\begin{equation}
\ba
j^{(0)}_y &\dps = e \text{Tr} \left(\frac{\hbar k_y}{m^*} \openone . n_{L,n}\right) = \frac{e^2B_z}{4\pi}\int kdk \Big\{\frac{2\hbar \tau_p k}{m^{*2}}n^{(0)}_k\Big\}=\frac{e^3 B_z \tau^2_p p}{m^{*2}}E_x,\\[3ex]
R_H^{(0)} &\dps = \lim\limits_{B_z \rightarrow 0} \frac{\sigma^{(0)}_{yx}}{B_z (\sigma^2_{xx} + \sigma^2_{xy})}=\frac{\frac{e^3 B_z \tau^2_p p}{m^{*2}}}{B_z \Big[\frac{e^2\tau_p}{2\pi m^*}k^2_{\text{F}}\Big]^2
}=\frac{1}{pe}.
\ea
\end{equation}

\section{Cubic terms}

\begin{figure}
	\includegraphics[scale=1]{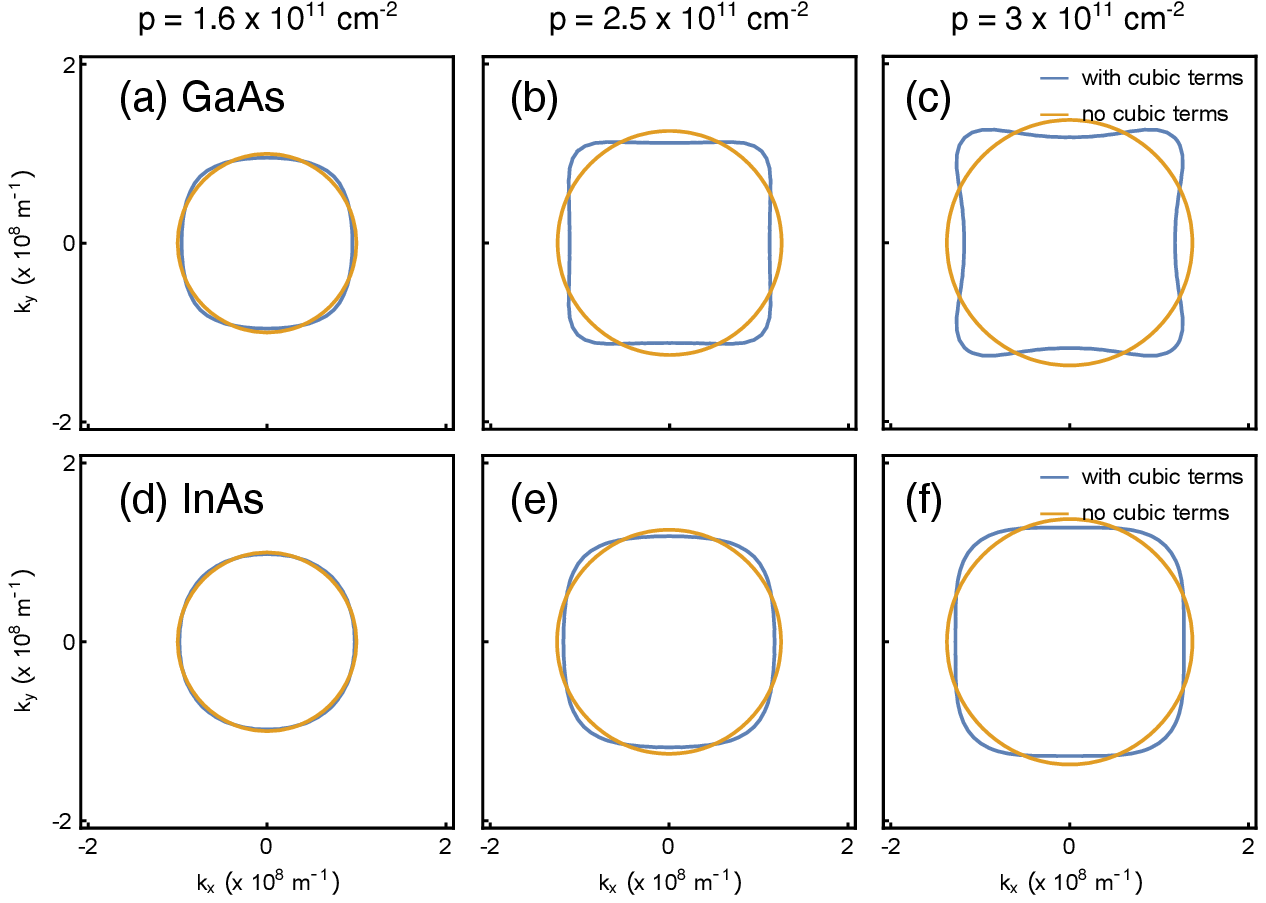}
	\caption{\label{fig:cubic_terms}Fermi contour with the cubic terms for (a), (d) $p = 1.59 \times 10^{11}$ cm$^{-2}$, (b), (e) $p = 2.5 \times 10^{11}$ cm$^{-2}$, and (c), (f) $p = 3 \times 10^{11}$ cm$^{-2}$ for GaAs and InAs, calculated using the Luttinger Hamiltonian, for a quantum well width $d = 20$ nm, and $B_{\parallel} = 0$ T.}
\end{figure}

Here we consider the cubic symmetry terms in the Luttinger Hamiltonian (i.e. terms  $ \propto (\gamma_3 - \gamma_2)$) (Fig.~\ref{fig:cubic_terms}). At a typical experimental density ($p = 1.59\times 10^{11}$ cm$^{-2}$, corresponding to $k_F = 1 \times 10^8 $ m$^{-1}$), the cubic terms are negligible. When the density is larger, the cubic terms become more prominent so that the Fermi contour are no longer isotropic. 
